\documentclass[letterpaper, 10 pt, conference]{ieeeconf}  

\IEEEoverridecommandlockouts                              
\usepackage[utf8]{inputenc}
\usepackage[T1]{fontenc}

\usepackage{amsmath,amssymb,amsfonts}
\usepackage{bm,bbm,siunitx,hyperref,graphicx}
\usepackage{float}	
\usepackage{xcolor}

\let\labelindent\relax

\usepackage{enumitem}
\usepackage{cite}
\usepackage{balance}

\title{\LARGE \bf
Data-Driven Strictly Positive Real System Identification with prior System Knowledge
}

\author{Nikhil Potu Surya Prakash$^{1}$ Zhi Chen$^{2}$ and Roberto Horowitz$^{3}$
\thanks{$^{1}$Nikhil Potu Surya Prakash is with Department of Mechanical Engineering,
        University of California, Berkeley, CA 94720, USA
        {\tt\small nikhilps@berkeley.edu}}%
\thanks{$^{2}$Zhi Chen is with Department of Mechanical Engineering,
        University of California, Berkeley, CA 94720, USA
        {\tt\small chenzhi@berkeley.edu}}%
\thanks{$^{3}$Roberto Horowitz is with Department of Mechanical Engineering,
        University of California, Berkeley, CA 94720, USA
        {\tt\small horowitz@berkeley.edu}}%
}

\begin{document}

\maketitle
\thispagestyle{empty}
\pagestyle{empty}

\begin{abstract}
Strictly Positive Real (SPR) transfer functions arise in many areas of engineering like passivity theory in circuit analysis and adaptive control to name a few. In many physical systems, it is possible to conclude that the system is Positive Real (PR) or SPR but system identification algorithms might produce estimates which are not SPR. In this paper, an algorithm to approximate frequency response data with SPR transfer functions using Generalized Orthonormal Basis Functions (GOBFs) is presented. Prior knowledge of the system helps us to get approximate pole locations, which can then be used to construct GOBFs. Next, a convex optimization problem will be formulated to obtain an estimate of the SPR transfer function. 

\end{abstract}

\section{Introduction}\label{sec:Introduction}
Strictly Positive Real (SPR) transfer functions arise in many areas of engineering like passivity theory in circuit analysis and adaptive control. In passivity theory, storage functions and lyapunov functions are related through the positive real lemma. Passivity theory and Hyper-stability are also used to analyze the stability of parameter adaptation algorithms which involves an SPR condition. Though the series-parallel adaptation algorithms (also called equation error methods)  automatically satisfy the SPR condition, they produce biased estimates of the plant in the presence of measurement noises. Parallel predictors (also called output error methods) prove to be better in the presence of measurement noises but suffer from stability issues. The stability condition in this case requires the transfer function with just the poles of the system being identified to be SPR which is not necessary to happen. To overcome this issue, a prefilter is designed such that the system formed by the ratio of the prefilter and the poles of the system being identified satisfies an SPR condition. But coming up with this filter has mostly been heuristic based on the prior knowledge of the location of the poles which is still not easy. This forms one of the motivating factors to come up with an algorithm for SPR system identification.

Frequency response measurements can be obtained for the plants of interest in many cases at least when the plants are stable and can be used to build plant models. The plant models can be used as initial conditions for the adaptation algorithms. But even with a precise knowledge of the frequency domain data, coming up with good estimates of the plants is hard as in some cases even when it is known that the plant is stable, the least squares fits might produce an estimate which is unstable. A nonconvex optimization problem is posed and solved in \cite{c2} to obtain stable estimates from frequency response data. System identification algorithms in the time domain like the series-parallel adaptation algorithm usually do not suffer from this issue as the error signals at every time instant are compared and the system usually converges to a stable model when the plant being identified is stable. Eignesystem Realization Algorithm (ERA) in \cite{c22} is an algorithm which estimates the system matrices by exciting the system with impulses. An extension of this is the Observer/Kalman filter identification (OKID) method in \cite{c1,c3,c4,c5,c6,c7,c24} which uses non-impulsive inputs and identifies the optimal Kalman fiter gains in the presence of noises with unknown variances and co-variances. These unique properties of time domain and frequency domain system identification techniques can be used in tandem to obtain SPR estimates.

In this paper, we will present a systematic methodology to obtain SPR estimates by combining the time domain response data and frequency response measurements of the system of interest. The time domain data can first be used to obtain an initial estimate of the system using OKID/ERA or series parallel adaptation algorithm etc. This estimated system is not necessarily SPR but the poles and zeros would be close to the real values. In the second step, the poles of the estimated system will be used to construct GOBFs. The actual system is assumed to be a linear combination of these GOBFs and their coefficients will be found such that $\mathcal{H}_\infty$ norm of the error between the actual frequency response measurements and the frequency response of the estimated system is minimized. The SPR condition on the system will be converted to a finite set of linear inequality constraints obtained by evaluating the SPR condition at various frequencies which are then incorporated into the optimization problem. As this is a data-driven technique, the SPR constraint is enforced only at a finite set of frequencies, the confidence level of the algorithm can be improved by using a large set of frequency response data. 

\section{Observer/Kalman filter Identification and Eigensystem realization Algorithm (OKID/ERA)}
In this section, a brief summary of OKID/ERA will be presented which could be used as the first step in obtaining an initial estimate of the plant from the time domain response of the plant. This initial estimate could be used to represent prior knowledge of the system in cases where there is absolutely no knowledge of the system being identified.

OKID/ERA is a framework for system identification when the process and measurement noises are assumed to be zero mean white noises with unknown variances and covariances. \cite{c24} extends this framework to a case when the noises are colored by an unknown stable process which is an important problem in hard disk drives as the runout is a colored noise. Any other system identification technique can also be used for obtaining the initial estimate like a series-parallel adaptation algorithm on the time domain signals or a simple Least-Squares algorithm on the frequency response data. 

Consider the following dynamical system in the state space form
\begin{align}\label{eq:augmentedstatespace}
\begin{split}
       x(k+1)&=Ax(k)+Bu(k)+w_p(k) \\
         y(k) &= Cx(k)+Du(k)+w_m(k) 
\end{split}
\end{align}
where $(A,B,C,D)$ are the system's state, input, output and feedthrough matrices respectively. $x(k)$, $y(k)$ and $u(k)$ are the system's state, measured output and input at the $k^{th}$ time instant. $w_p(k)$ and $w_m(k)$ are the process and measurement noises at the $k^{th}$ time instant. $w_p$ and $w_m$ are assumed to be zero mean white noises with unknown variance and co-variance.

The pair $(A,C)$ is assumed to be observable, and hence a steady state Kalman filter can be designed for the system if we exactly know the matrices $(A,B,C,D)$. Hence a Kalman filter gain $K$ exists such that $A-KC$ is Schur. With the filter gains defined (as unknowns), the observer dynamics can be expressed as
\begin{equation}\label{observer}
    \begin{split}
        \hat{x}(k+1) &= (A-KC)\hat{x}(k)+(B-KD)u(k)+Ky(k) \\
        \hat{y}(k) &= C\hat{x}(k)+Du(k)
    \end{split}
\end{equation}
Let $F = A-KC,\; H = B-KD$ and $G = K$\\
Defining $L = [H\;G]$ and $\nu_x(k) = [u^T(k)\;y^T(k)]^T$ for brevity, we get the observer in predictor form as
\begin{equation}\label{FL}
    \begin{split}
        \hat{x}(k+1) = F\hat{x}(k)+Lv_x(k) \\
        \hat{y}(k) = C\hat{x}(k)+Du(k)
    \end{split}
\end{equation}
Using \eqref{observer}, the observer state at $k^{th}$ time instant for any $p$ can be derived as
\begin{equation}\label{obsstateprg}
    \hat{x}(k) = F^p \hat{x}(k-p)+Tz(k)
\end{equation}

where 
\begin{align*}
    T &= 
    \begin{bmatrix}
    I & F & ... & F^{p-2} & F^{p-1}
    \end{bmatrix} L, \; \\
    z(k) &= 
    \begin{bmatrix}
    \nu^T_x(k-1) & \nu^T_x(k-2) &...& \nu^T_x(k-p)
    \end{bmatrix}
\end{align*}
The stability of the observer ensures that $F^p$ becomes negligible for sufficiently large values of  $p \; (p>>n)$ and hence the observer state in \eqref{obsstateprg} becomes
\begin{equation}\label{obsaprxstate}
    \hat{x}(k) \approx Tz(k)
\end{equation}
Substituting \eqref{obsaprxstate} in \eqref{FL}, we get the estimated output as
\begin{equation}\label{estmy}
    \hat{y}(k) = CTz(k)+Du(k)
\end{equation}
The estimated output in \eqref{estmy} is related to the measured output as
\begin{equation}
    y(k) = \Phi \nu(k)+\epsilon(k)
\end{equation}
where $\Phi = [D\;CL\;CFL\;...\;CF^{p-2}L\;CF^{p-1}L]$, $\nu(k) = [u^T(k) z^T(k)]^T$ and $\epsilon(k)$ is the error between the measured output and the estimated output.\\
The outputs at different time instants can be collected and stacked to obtain the following equation
\begin{equation}\label{Ystack}
    Y = \Phi V+E
\end{equation}
where
\begin{subequations}
    \begin{equation}
        Y = [y(p)\;y(p+1)\;...\;y(l-1)]
    \end{equation}
    \begin{equation}
        V = [\nu(p)\;\nu(p+1)\;...\;\nu(l-1)]
    \end{equation}
    \begin{equation}
        E = [\epsilon(p)\;\epsilon(p+1)\;...\;\epsilon(l-1)]
    \end{equation}
\end{subequations}
for $l$ measurements. The best estimate of $\Phi$ is obtained using least squares formulation as $\hat{\Phi} = YV^{\dagger}$ ($\dagger$ denotes the pseudo inverse). Various system matrices and the Kalman filter gain can be extracted from $\hat{\Phi}$ using the Eigensystem Realization Algorithm (ERA).

Eigensystem Realization Algorithm, first developed in \cite{c22}, is a system identification technique used most popularly for aerospace and civil structures from the input and output time domain data. Though ERA uses impulses as inputs to excite the system, it can be intertwined with OKID framework to estimate the system matrices using non-impulsive inputs. There are many variants to the ERA depending on the application and the type of data collected. In this section we will summarize one of the variants to extract estimates of the system matrices and the Kalman filter gain from $\hat{\Phi}$. 

It can be easily seen that $D$ is the first column in $\hat{\Phi}$. From the Markov parameters in $\hat{\Phi}$, the following Hankel matrices can be defined
\begin{equation}
\mathcal{H}(0) =
    \begin{bmatrix}
    CL & CFL & CF^2L & \dots\\
    CFL & CF^2L &CF^3L & \dots \\
    \vdots & \vdots &\ddots
    \end{bmatrix} = \mathcal{O}\mathcal{C}
\end{equation}

\begin{equation}
\mathcal{H}(1) = 
    \begin{bmatrix}
    CFL & CF^2L & CF^3L & \dots\\
    CF^2L & CF32L &CF^4L & \dots \\
    \vdots & \vdots &\ddots
    \end{bmatrix} = \mathcal{O}F\mathcal{C}
\end{equation}
where $\mathcal{O}$ is the observability matrix and $\mathcal{C}$ is the controllability matrix of the dynamical system in \eqref{FL} with $\nu_x$ as its input.\\
Using singular value decomposition, $\mathcal{H}(0)$ can be decomposed as
\begin{equation}
    \mathcal{H}(0) = U \Sigma V^T = 
    \begin{bmatrix}
          U_{+} & U_{-}
    \end{bmatrix}
    \begin{bmatrix}
          \Sigma_{+} & 0 \\
          0 & \Sigma_{-}
    \end{bmatrix}
    \begin{bmatrix}
          V^T_{+} \\ V^T_{-}
    \end{bmatrix}
\end{equation}
The subscript '$+$' denotes the singular values above a specified threshold and '$-$' for the singular values below the threshold. The desired degree of the estimated system can be decided from these dominant singular values.\\
The Observability and Controllability matrices can be split as follows (\cite{c23} presents other ways in which the matrices can be split).
\begin{equation}
\mathcal{O} = U_{+}\Sigma_{+}^{\frac{1}{2}} 
\end{equation}
\begin{equation}
\mathcal{C} = \Sigma_{+}^{\frac{1}{2}}V^T_{+} 
\end{equation}
Now using the Hankel matrix $\mathcal{H}(1)$, the matrix $F$ can be obtained as
\begin{align}
\mathcal{H}(1) &= \mathcal{O}F\mathcal{C} = U_{+}\Sigma_{+}^{\frac{1}{2}}F\Sigma_{+}^{\frac{1}{2}}V^T_{+} \nonumber \\ \implies F &=  \Sigma_{+}^{-\frac{1}{2}}U^T_{+}\mathcal{H}(1)V_{+}\Sigma_{+}^{-\frac{1}{2}}
\end{align}
Since $\mathcal{C}$ is the controllability matrix of \eqref{FL}, $L$ can be obtained from the first few columns of $\mathcal{C}$ and similarly $C$ can be obtained from the first few rows of $\mathcal{O}$ based on the dimensions. Further $H$ and $G$ can be obtained from $L$ as $L=[H\;G]$.
The Kalman filter gain $K=G$. Now, the state matrix $A$ and the input matrix $B$ can be estimated from $F$ and $H$ matrices respectively as follows\\
\begin{equation}
    F=A-KC \implies A=F+GC
\end{equation}
\begin{equation}
    H=B-KD \implies B=H+GD
\end{equation}

\section{Generalized Orthonormal Basis Functions (GOBF)}
In this section a brief description of Generalized Orthonormal Basis Functions (GOBFs) will be provided.
Orthogonal functions were used in \cite{c27,c28} for the identification of a finite
sequence of expansion coefficients. The idea is based on the fact that every stable system can be represented as a unique series expansion in terms of a prechosen basis. A representation with finite number of terms can serve as an approximate model. Historically, the coefficients of the series expansion have been estimated from the input-output time domain data, but they can also be estimated from the frequency domain data as will be shown later.

Consider a stable discrete time system $G(q)$. Then there exists a unique series expansion in terms of orthonormal basis function $\{ f_k(q) \}_{k = 1,2,\dots}$ such that 
\begin{equation}
    G(q) = \sum_{k=1}^\infty a_k f_k(q)
\end{equation}
where $\{a_k\}_{k = 1,2,\dots}$ are real coefficients. Since $\{ f_k(q) \}_{k = 1,2,\dots}$ are orthonormal, they satisfy the following property
\begin{equation}
    \frac{1}{2\pi}\int_{-\frac{\pi}{T_s}}^{\frac{\pi}{T_s}}f_l(e^{j\omega T_s})f_m(e^{-j\omega T_s})d\omega = \bigg\{
    \begin{matrix}
    1 & if & l=m \\
    0 & if & l\neq m
    \end{matrix}
\end{equation}
If we are choosing a finite number of basis functions, it is profitable to use filters that reflect the dominant dynamics of the system i.e., the poles. The poles of the estimated system from OKID/ERA will be used to build these filters. The most well known filters and Laguerre filters for well damped poles and Kautz filters for underdamped poles. \\\\
\textit{Laguerre filters:}
\begin{equation*}
    L_k(q,a) = \frac{\sqrt{1-a^2}}{q-a}\bigg[\frac{1-aq}{q-a}\bigg]^{k-1}, \;  a\in \mathbb{R},|a| < 1, k \geq 1
\end{equation*}
\textit{Kautz filters:} \\
for odd $k$
\begin{equation*}
    \Psi_k(q,a) =  \frac{\sqrt{1-c^2}(q-b)}{q^2+b(c-1)q-c}\bigg[\frac{-cq^2+b(c-1)q+1}{q^2+b(c-1)q-c}\bigg]^{\frac{k-1}{2}}
\end{equation*}
for even $k$
\begin{equation*}
    \Psi_k(q,a) =  \frac{\sqrt{(1-c^2)(1-b^2)}}{q^2+b(c-1)q-c}\bigg[\frac{-cq^2+b(c-1)q+1}{q^2+b(c-1)q-c}\bigg]^{\frac{k-2}{2}}
\end{equation*}
\begin{equation*}
b,c \in \mathbb{R}, |b| < 1 , |c| <1 , k\geq 1
\end{equation*}
Here $a,b$ and $c$ are the parameters that decide the location of the poles.

\section{SPR System Estimation}
In this section, utilizing the GOBFs, identification of the SPR transfer function will be posed as an optimization problem. Various texts provide various definitions of SPR transfer functions, so, to be consistent, we present the definition from \cite{c26}. \\\\
\textbf{Definition:} A discrete time transfer function $G(q)$ is said to be strictly positive real if
\begin{enumerate}
    \item $G(q)$ is Schur
    \item $Re[G(e^{j\omega T_s})]>0 \; \forall \; \omega \in [-\frac{\pi}{T_s},\frac{\pi}{T_s}]$  
\end{enumerate}
Note that in discrete time, a transfer function can be SPR only if its relative degree is zero.\\ The first condition is automatically satisfied when we choose the GOBFs to have stable poles. After choosing the GOBFs, we proceed with formulating the identification problem as a convex optimization problem. The transfer function estimate can be approximated as a linear combination of the GOBFs as 
\begin{equation}\label{eq:ghat}
    \hat{G}(q) = \sum_{i=1}^{N} a_i f_i(q) = \theta^T\Phi(q)
\end{equation}
where $f_i(q) $ is either a Laguerre or a Kautz filter and the constants $a_i$ are to be determined. Let $G(e^{j\omega_n T_s})$ denote the frequency response measurements of the system at $p$ different frequencies $\omega_n$ for $n = 1,...,p$
\begin{equation}
    \hat{G}(e^{j\omega_n T_s}) = \sum_{i=1}^{N} a_i f_i(e^{j\omega_n T_s}) = \theta^T\Phi(e^{j\omega_n T_s})
\end{equation}
where $\theta = [a_1 \; a_2 \; \dots \; a_N]^T$ and $\Phi(q) = [f_1(q) \; f_2(q) \; \dots \; f_n(q)]^T$. \\
Since $\hat{G}$ must be strictly positive real, $Re(\hat{G}) \geq \epsilon$ for some $\epsilon >0 $ for some user defined tolerance $\epsilon$ i.e.,
\begin{align}
&\hat{G}(e^{j\omega_n T_s})+\hat{G}(e^{-j\omega_n T_s}) \geq \epsilon \nonumber \\
\implies 
&\sum_{i=1}^{N} a_i Re(f_i(e^{j\omega_n T_s})) \geq \epsilon \nonumber \\
\implies 
&\theta^T Re(\Phi(e^{j\omega_n T_s})) \geq \epsilon
\end{align}
which forms a set of linear inequality constraints.\\
The problem of finding the constants $a_i$ can now be formulated as a convex quadratic program by minimizing the squared weighted/unweighted $\mathcal{H}_{\infty}$ norm of the difference between the estimate's frequency response and the frequency response of the actual plant with the SPR constraints formulated as linear constraints as follows
\begin{equation}
\begin{aligned}
\hat{G}^* = & \underset{\hat{G}}{\text{argmin}}
& & \sum_{n=1}^{p} ||G(e^{j\omega_n T_s})-\hat{G}(e^{j\omega_n T_s})||^2_2\\
& \text{subject to}
& & Re(\hat{G}(e^{j\omega_n T_s})) \geq \epsilon, \; n = 1, \ldots, p.
\end{aligned}
\end{equation}
which from \eqref{eq:ghat} is equivalent to 
\begin{equation}
\begin{aligned}
\theta^* = & \underset{\theta}{\text{argmin}}
& & \sum_{n=1}^{p} ||G(e^{j\omega_n T_s})-\theta^T\Phi(j\omega_n T_s)||^2_2\\
& \text{subject to}
& & \theta^T Re(-\Phi(e^{j\omega_n T_s})) \leq -\epsilon, \; n = 1, \ldots, p.
\end{aligned}
\end{equation}
The same procedure can be used to come up with estimates that satisfy the SPR condition on the ratio of the real transfer function and its estimate to ensure stability in adaptive control algorithms. In case of a constraint like $Re(\frac{\hat{G}(e^{j\omega T_s})}{G(e^{j\omega T_s})})>0$, we can formulate the optimization problem as follows 
\begin{equation}
\begin{aligned}
\hat{G}^* = & \underset{\hat{G}}{\text{argmin}}
& & \sum_{n=1}^{p} ||G(e^{j\omega_n T_s})-\hat{G}(e^{j\omega_n T_s})||^2_2\\
& \text{subject to}
& & Re(\frac{\hat{G}(e^{j\omega_n T_s})}{G(e^{j\omega_n T_s})}) \geq \epsilon, \; n = 1, \ldots, p.
\end{aligned}
\end{equation}
\section{simulations}
In this section, the SPR identification algorithm presented in the previous section will be demonstrated on two simple transfer functions. The first example considers the case when the actual transfer function is SPR and the second example considers the case in which the actual transfer function is not SPR. In both the cases, the algorithm will be able to find a transfer function that is SPR whose frequency response is as close as possible to the frequency response of the actual plant. \\

\begin{figure}[htpb]
    \centering
    \includegraphics[width=\columnwidth]{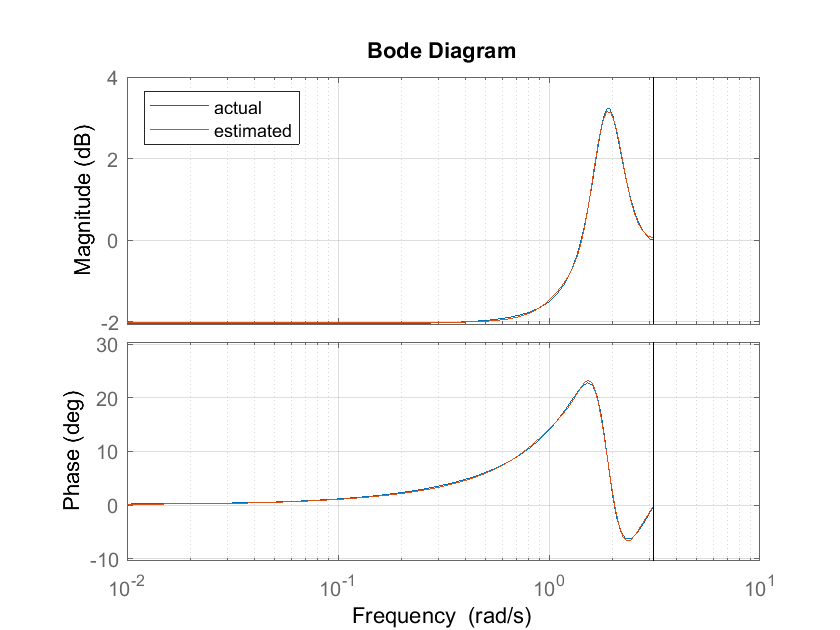}
    \caption{Frequency responses of the actual plant and the estimated plant}
    \label{fig:bode}
\end{figure}

\begin{figure}[htpb]
    \centering
    \includegraphics[width=0.9\columnwidth]{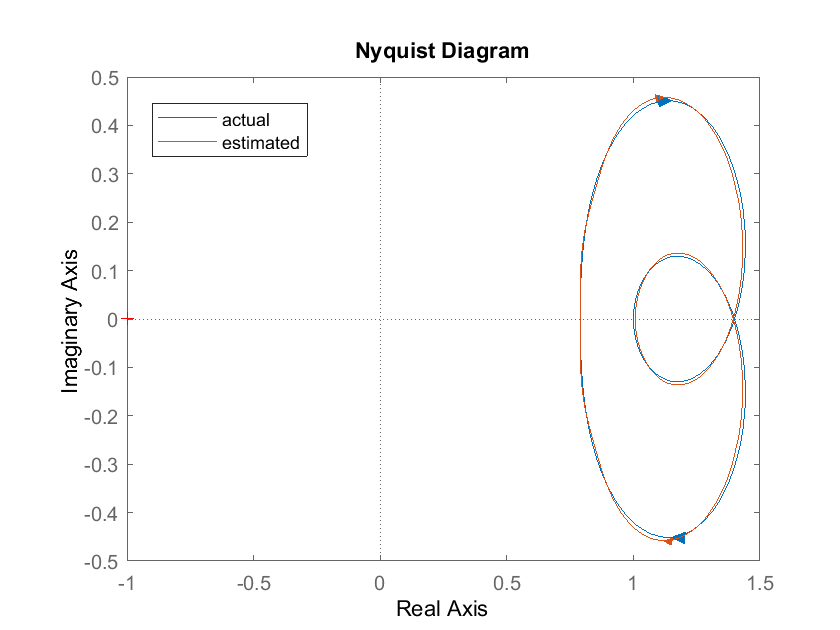}
    \caption{Nyquist plot of the actual plant and the estimated plant}
    \label{fig:nyquist}
\end{figure}

\textit{Example 1:}
In this example, a simple second order discrete time transfer function will be considered which is strictly positive real. The transfer function corresponding to the frequency response considered is 
\begin{equation*}
    G(q) = \frac{q^2+0.2q+0.3}{q^2+0.4q+0.5}
\end{equation*}
Kautz filters were built with c = -0.2 and b = -0.33. It can be seen that the values used for b and c do not represent the actual poles of the system and yet the simulation results will show that a good fit can be obtained. N was chosen to be 8 for the algorithm. The Bode plot in fig.\ref{fig:bode} and Nyquist plot in fig.\ref{fig:nyquist} show a close fit and that the SPR conditions are also being satisfied.

\textit{Example 2:}
In this example, a slight variation of the first example is considered which makes the Nyquist plot extending to the left half of the imaginary axis. The transfer function corresponding to the frequency response considered is 
\begin{equation*}
    H(q) = \frac{0.25q^2+0.2q+0.3}{q^2+0.4q+0.5}
\end{equation*}
Kautz filters were chosen to be the basis functions with the same parameters as in example 1. In this case, the Bode plot in fig.\ref{fig:bode} shows a fit that deviates significantly from the original frequency response. Yet, the Nyquist plot in fig.\ref{fig:nyquist} shows a fit that satisfies the SPR conditions.
\begin{figure}[htpb]
    \centering
    \includegraphics[width=\columnwidth]{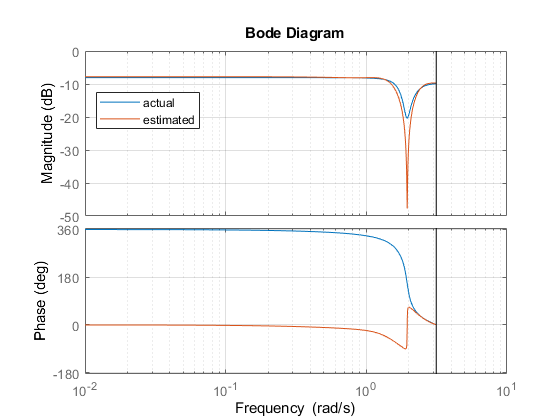}
    \caption{Frequency responses of the actual plant and the estimated plant}
    \label{fig:bode2}
\end{figure}
\begin{figure}[htpb]
    \centering
    \includegraphics[width=0.9\columnwidth]{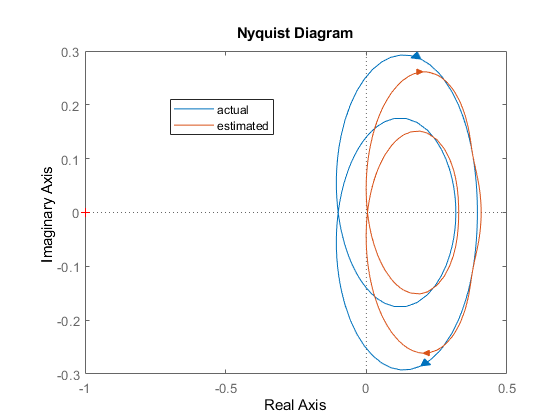}
    \caption{Nyquist plot of the actual plant and the estimated plant}
    \label{fig:nyquist2}
\end{figure}
\section{Conclusions}\label{sec:Conc}
In this paper a data-driven technique for estimating strictly positive real transfer functions from its frequency response with prior knowledge of the approximate location of poles is presented. The problem is posed as a convex optimization problem with a quadratic objective function and linear constraints. The robustness of the algorithm was demonstrated through examples in which the real poles were far from the poles considered to build the basis functions.
\section*{ACKNOWLEDGMENTS}
We would like to acknowledge the feedback and financial support from ASRC hosted by the International Disk Drive Equipment and Materials Association (IDEMA).


\begin{thebibliography}{99}

\bibitem{c1} Jer-Nan Juang, Minh Phan, Lucas G. Horta and Richard W. Longman, "identification of Observer/Kalman Filter Markov Parameters: Theory and Experiments", in Journal of Guidance Control and Dynamics, vol. 16, No.2,March-April 1993.
\bibitem{c2} W.-K. Chen, Linear Networks and Systems (Book style).	Belmont, CA: Wadsworth, 1993, pp. 123--135.
\bibitem{c3} Francesco Vicario, Minh Q Phan, Raimondo Betti, Richard W Longman, "OKID via Output Residuals: A Converter from Stochastic to Deterministic System Identification", in Journal of Guidance, Control, and Dynamics, vol. 40, Issue 12, 2007, pp. 3226--3238.
\bibitem{c4} Guojun Sun, Richard W Longman, Raimondo Betti, Zhihua Chen, Suduo Xue, "Observer Kalman filter identification of suspen-dome", in Journal of Mathematical Problems in Engineering, 2017.
\bibitem{c5} Francesco Vicario, Minh Q Phan, Richard W Longman, Raimondo Betti, "Generalized framework of OKID for linear state-space model identification", in Modeling, Simulation and Optimization of Complex Processes HPSC 2015, Springer, Cham, 2017, pp. 249--260.
\bibitem{c6} Francesco Vicario, Minh Q Phan, Raimondo Betti, Richard W Longman, "Output‐only observer/Kalman filter identification $O^3KID$", in Journal of Structural Control and Health Monitoring, Vol. 22, Issue 5, 2015, pp. 847--872.
\bibitem{c7} Francesco Vicario, Minh Q Phan, Raimondo Betti, Richard W Longman, "Extension of OKID to Output-Only System Identification", 2014.
\bibitem{c8}Phan, M. Q., Vicario, F., Longman, R. W., and Betti, R. (November 8, 2017). "State-Space Model and Kalman Filter Gain Identification by a Kalman Filter of a Kalman Filter." ASME. Journal of Dynamical Systems, Measurement, and Control, March 2018; 140(3): 030902. https://doi.org/10.1115/1.4037778
\bibitem{c9} Lennart Ljung, Keith Glover, "Frequency domain versus time domain methods in system identification", in Automatica, Vol. 17, Issue 1, 1981, pp. 71-86.
\bibitem{c10} MATLAB. version 9.10.0 (R2016b). Natick, Massachusetts: The MathWorks Inc.,2016.
\bibitem{c11} Zhang, Qinghua. “Using Wavelet Network in Nonparametric Estimation”, IEEE Transactions on Neural Networks 8, no. 2 , March 1997. https://doi.org/10.1109/72.557660.
\bibitem{c12} Sjöberg, Jonas, Qinghua Zhang, Lennart Ljung, Albert Benveniste, Bernard Delyon, Pierre-Yves Glorennec, Håkan Hjalmarsson, and Anatoli Juditsky. “Nonlinear Black-Box Modeling in System Identification: A Unified Overview”, Automatica 31, no. 12 (December 1995): 1691–-1724. https://doi.org/10.1016/0005-1098(95)00120-8.
\bibitem{c13} Ljung, Lennart, and Torkel Glad. Modeling of Dynamic Systems. Prentice Hall Information and System Sciences Series. Englewood Cliffs, NJ: PTR Prentice Hall, 1994.
\bibitem{c14} Ljung, Lennart. System Identification: Theory for the User. Second edition. Prentice Hall Information and System Sciences Series. Upper Saddle River, NJ: PTR Prentice Hall, 1999.
\bibitem{c15} Söderström, Torsten, and Petre Stoica. System Identification. Prentice Hall International Series in Systems and Control Engineering. New York: Prentice Hall, 1989.
\bibitem{c16} Lennart Ljung, "Analysis of recursive stochastic algorithms", IEEE transactions
on automatic control 22.4, 1977, pp. 551--575.
\bibitem{c17} Behrooz Shahsavari, Jinwen Pan, and Roberto Horowitz, "Adaptive rejection of periodic
disturbances acting on linear systems with unknown dynamics", IEEE
\bibitem{c18} Simon Haykin and Bernard Widrow, "Least-mean-square adaptive filters", Vol. 31. John Wiley \& Sons, 2003.
\bibitem{c19} Simon S Haykin, "Adaptive filter theory", Pearson Education India, 2008.
\bibitem{c20} Roberto Horowitz et al, "Dual-stage servo systems and vibration compensation in
computer hard disk drives", Control Engineering Practice 15.3 (2007), pp. 291--305.
\bibitem{c21}A. Tiano, R. Sutton, A. Lozowicki, W. Naeem,"Observer Kalman filter identification of an autonomous underwater vehicle", Control Engineering Practice, Vol. 15, Issue 6, 2007, pp. 727--739, ISSN 0967-0661, https://doi.org/10.1016/j.conengprac.2006.08.004.
\bibitem{c22} Juang, J.-N., Pappa, R. S., "An Eigensystem Realization Algorithm for Modal Parameter Identification and Model Reduction", Journal of Guidance, Control, and Dynamics, Vol. 8, Issue 5, pp.620–-627
\bibitem{c23} H.P.Gavin, "Eigensystem Realization", Course Notes, Duke University, April 15, 2020.
\bibitem{c24}Nikhil Potu Surya Prakash, Zhi Chen and Roberto Horowitz, "System Identification in Multi-Actuator Hard Disk Drives with Colored Noises using Observer/Kalman Filter Identification (OKID) Framework", 2021, arXiv:2109.12460
\bibitem{c25}Richard G. Hakvoort and Paul M. J. Van Den Hof (1994) Frequency domain curve fitting with maximum amplitude criterion and guaranteed stability, International Journal of Control, 60:5, 809-825, DOI: 10.1080/00207179408921496
\bibitem{c26}Hassan K. Khalil, "Nonlinear Control", Pearson Education, 2015, ISBN-10: 0-13-349926-X,ISBN-13:978-0-13-349926-1
\bibitem{c27}Peter Lindskog,"Algorithms and Tools for System Identification Using Prior Knowledge",PhD Thesis,1994, ISBN 91-7871-422-2.
\bibitem{c28}Bo Wahlberg, " System Identification Using Kautz Filters",IEEE Transactions on Automatic Control, Vol. 39, NO. 6, June 1994.

\end{thebibliography}
\end{document}